\documentclass[12pt]{iopart}
\usepackage{iopams}
\usepackage{harvard}
\usepackage{graphicx}
\usepackage{amstext}
\usepackage{hyperref}
\usepackage{textcomp}
\usepackage{upgreek}

\makeatletter
\newcommand{\seteqabctag}[1]{\def\@eqnnum{{\normalfont \normalcolor (\theequation #1)}}}
\makeatother

\begin{document}

\title[The normal field instability under side-wall effects]{The normal field instability under side-wall effects: comparison of experiments and computations}

\author{C Gollwitzer$^1$, A N Spyropoulos$^2$, 
A G Papathanasiou$^2$, A G Boudouvis$^2$, 
and R Richter$^1$}

\address{$^1$ Experimentalphysik V, University of Bayreuth, 95440 Bayreuth, Germany}
\address{$^2$ School of Chemical Engineering, National Technical University of Athens, 15780 Athens, Greece}
\ead{Christian.Gollwitzer@uni-bayreuth.de}

\begin{abstract}
We consider a single spike of ferrofluid, arising in a small cylindrical container, when a vertically oriented
magnetic field is applied. The height of the spike as well as the surface topography is measured experimentally
by two different technologies and calculated numerically using the finite element method.
As a consequence of the finite size of the container, the numerics uncovers an imperfect
bifurcation to a single spike solution, which is forward. This is in contrast to the
standard transcritical bifurcation to hexagons, common for rotational symmetric systems
with broken up-down symmetry. The numerical findings are corroborated in the experiments.
The small hysteresis observed is explained in terms of a hysteretic wetting of the
side wall.
\end{abstract}

\maketitle

\section{Introduction}
Rotational symmetric systems with broken up-down symmetry become first unstable due to a
transcritical bifurcation to hexagons, which is hysteretic ~\cite{cross1993}. Examples are
non-Boussinesq Rayleigh-B\`enard convection~\cite{busse1967} and chemical reactions of the
Turing type~\cite{turing1952}. The same is true, when a layer of magnetic fluid is exposed
to a normal magnetic field. Above a certain critical induction $B_c$, a hexagonal pattern of liquid spikes appears on the
surface of the fluid. This striking phenomenon was first reported by
\citeasnoun{cowley1967} and described in terms of a linear stability analysis. This
observation $40$ years ago triggered numerous efforts to describe also the
nonlinear aspects of the phenomenon theoretically. \citeasnoun{gailitis1977}, later
\citeasnoun{friedrichs2001} and \citeasnoun{friedrichs2002}
and most recently \citeasnoun{bohlius2006b} used the principle of free energy minimisation to
predict the pattern ordering, wavelength and final amplitude of the peaks on an infinitely
extended surface. The numerical computations by \citeasnoun{boudouvis1987phd} and
\citeasnoun{boudouvis1987} specify quantitatively the hysteresis in spike height in unbounded ferrofluid
pools. \citeasnoun{matthies2005} calculate also the dynamics of an infinite periodic lattice of peaks.

The experiments, however,
are performed with limited amounts of fluid. A finite
layer depth in the vertical dimension has been incorporated into the theory by
\citeasnoun{friedrichs2001}. According to \citeasnoun{lange2000}, 
the infinitely deep limit is well approximated if the depth exceeds at
least the wavelength of the pattern. However, in the horizontal dimension the
finite container size has not been considered. Therefore,
experimental realizations approximated this limit of an infinitely extended
layer by several different approaches.
\citeasnoun{abou2001} used
a very large aspect ratio, whereas \citeasnoun{gollwitzer2007} employed an
inclined container edge. \citeasnoun{richter2005}
used a magnetic ramp to minimize the influence of the border for the Rosensweig 
instability, whereas \citeasnoun{embs2007} independently applied it to the Faraday instability in ferrofluid.  

The question arises, what happens if the container size is intentionally reduced until
only a single spike is left. In this case, all symmetries are kept, nonetheless the
character of the bifurcation may change. Indeed, in experiments before the seminal
work of \citeasnoun{bacri1984}, it was difficult to uncover a hysteresis due to the small container size. 

Although there have been numerous experiments in small containers merely
because they are simple and cheap, a systematic study of the influence of the constrained
geometry on the bifurcation is missing. One reason is, that model descriptions which deal with
the finite container size are rare. So far, we know only the work by
\citeasnoun{friedrichs2000} where the free surface is modeled by a four
parameter function to fit the measurements by \citeasnoun{mahr1998}. In this case, a
highly susceptible fluid still showed a hysteretic transition. 

In the present article, we demonstrate, that a transcritical bifurcation to hexagons, as
found by~\citeasnoun{gollwitzer2007}, e.g.,  becomes an imperfect supercritical bifurcation to a
single spike if the container size is reduced sufficiently. This is the outcome of a
numerical model which is able to calculate the stable and unstable solutions for given container
size and fluid parameters.  It also takes into account the side-wall effects, namely the
wetting and the fringing field. 
We compare the numerical results with our measurements of
the surface topography. For the first time, we apply two different techniques which are capable of recording the 
amplitude~\cite{megalios2005} and also
the full topography of the fluid surface~\cite{richter2001}, to the same experiment. 

In the following two sections, we give an overview over the experimental
methods. Subsequently,  we
describe the numerical computations. Finally, we compare all three results.

\section{Measurements of the material properties}
\begin{table}
\centering
\begin{minipage}{0.8\linewidth}
\begin{tabular}{lcrl}
Quantity & & Value \phantom{$\pm$}& Error\\[3pt]
Surface tension\footnote{The absolute error of the measurement is unknown. The
error given here is taken from the analysis by \citeasnoun{harkins1930}} 
&$\sigma$ & $30.57\ \pm$ & $0.1\,\mathrm{mN\,m^{-1}}$\\
Density			&$\rho$ &  $1236\ \pm$ & $1\,\mathrm{kg\,m^{-3}}$\\
Contact angle with the container wall & $\theta_c$ & $10\pm0.3$\,\textdegree\\
Viscosity		&$\eta$  & $120\ \pm$ & $5\,\mathrm{mPa\,s}$\\
Saturation magnetization &$M_S$ &  $28.7\ \pm$ & $0.1\,\mathrm{kA\,m^{-1}}$ \\
Initial susceptibility & $\chi_0$ & $1.2023\ \pm$ & $0.005$
\end{tabular}
\caption{Material properties of the ferrofluid  APG\,512a (Lot~F083094CX) from Ferrotec Co.}
\label{tbl:props}
\end{minipage}
\end{table}

We used the ferrofluid APG\,512a (Lot~F083094CX) from Ferrotec for all
experiments. It is based on an ester with a very low vapor pressure,
suitable for vacuum pumps. It has an excellent long-term stability. Over one
year, the critical induction has not changed by more than $3\,\%$. In contrast
to less stable magnetic liquids, the formation of agglomerates in the tips of
the Rosensweig spikes was not observed. After applying magnetic fields for an
hour, the field was switched off. Neither the visual inspection nor the X-ray
images unveiled any agglomerates at the site of the spikes.

The Rosensweig instability is a counterplay between gravitational and surface terms 
on the one hand and magnetic forces on the other hand~\cite{cowley1967}.
Therefore, a set of basic material properties of the fluid is necessary for a
comparison with the theory, namely the surface tension $\sigma$, the density
$\rho$ and the magnetization curve $M(H)$. These quantities are summarized in
table~\ref{tbl:props}.

The surface tension has
been measured using a commercial ring tensiometer (LAUDA TE1).
This device wets a ring made from platinum wire, pulls
it off the fluid surface and determines the maximum force acting on the ring,
from which the surface tension can be computed following
\citeasnoun{dunouy1919}. According to an analyis by \citeasnoun{harkins1930}, the error for
this method is smaller than $0.25\,\%$, given that the density of the fluid
and the geometry of the ring are known with sufficient accuracy. 

The density $\rho$ has been measured using a commercial vibrating-tube densimeter (DMA
4100 by Anton Paar). This device enables us to determine the density with an
error of $0.01\,\%$.  

The contact angle $\theta_c$ was determined with the contact angle system OCA~20
(Dataphysics) by optical means. Three measurements were performed at the inner side wall of the container,
which was tilted by $90\,$\textdegree. The difference between advancing and receding angle
could not be measured in this way.  

\begin{figure}
\centering
\includegraphics[width=0.7\linewidth]{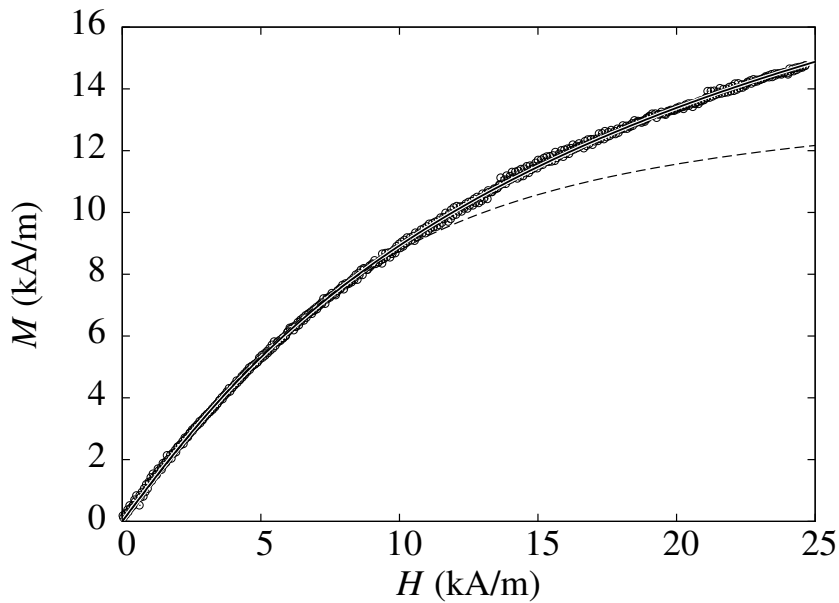}
\caption{The magnetization curve of the ferrofluid. The circles denote the
experimental values, the solid line is a fit with the model by
\protect\citeasnoun{ivanov2001}. The dashed line is a fit with the Langevin function,
that is valid only up to $H\lessapprox 10\,\mathrm{kA/m}$. } 
\label{fig:magcurve}
\end{figure}
The magnetization curve $M(H)$ of the ferrofluid has the biggest influence on the
surface pattern. It has been meticulously measured using a fluxmetric magnetometer
consisting of a Helmholtz pair of sensing coils with $6800$ windings and a
commercial integrator (Lakeshore Fluxmeter 480). The sample has been held in a
spherical cavity with a diameter of $12.4\,\mathrm{mm}$. This spherical shape
ensures a homogeneous magnetic field inside the sample and an exact homogeneous
demagnetization factor of $1/3$, which makes it possible to get accurate results
over the whole range of $H$.  Figure~\ref{fig:magcurve} shows the magnetization curve
of our ferrofluid. The solid and the dashed line provide two analytic
approximations to $M(H)$. The dashed line is a fit with the Langevin
approximation for monodisperse colloidal suspensions and provides the
constitutive equation~\cite{rosensweig1985} 
\begin{equation}
\left|\underline M\right| = p \left[ \coth\left(\tau \left|\underline
H\right|\right) - \frac{1}{\tau \left|\underline H\right|}\right].
\label{eq:langevin}
\end{equation}
Only the data points within a range of $H\in [0\dots 10\,\mathrm{kA/m}]$ have
been taken into account for the estimation of the adjustable parameters
$p=14.6\,\mathrm{kA/m}$ and
$\tau=0.24\,\mathrm{m/kA}$. A satisfying fit of the whole curve with this equation is not
possible, because real ferrofluids consist of magnetic particles with a broad
size distribution~\cite{popplewell1995b}. We therefore make use of a model for dense
polydisperse magnetic fluids put forward by \citeasnoun{ivanov2001}. The solid line in
figure~\ref{fig:magcurve} displays the best fit with that model. 
The saturation magnetization given in table~\ref{tbl:props} is extrapolated from there. 
This extrapolation indicates together with the manufacturer
information $M_S\approx26\,$kA/m$\pm 10\,\%$, that this model is very well
fitted to our dense ferrofluid~\citeaffixed{ivanov2007}{cf.}.

\section{Measurements of the surface pattern}
We fill a cylindrical container, machined from aluminum, with the ferrofluid and
expose it to a magnetic induction ranging from
$B=7.6\,\mathrm{mT}$ to $B=37.7\,\mathrm{mT}$. The depth of the container
amounts to $20\,\mathrm{mm}$ and the diameter is $29.7\,\mathrm{mm}$. This diameter is
chosen by trial such that only one single spike emerges in the centre of the
vessel for all magnetic inductions we apply. 
From the weight of the filled container and the density, we
calculate the amount of ferrofluid filled into the container to $V_0 =
6.387\,\mathrm{ml}$, which is equivalent to a filling height of
$D=9.22\,\mathrm{mm}$.
Two complementary experimental methods were used to determine the height of the
emerging spike in the centre of the vessel: the X-ray method by
\citeasnoun{richter2001} and the laser method by \citeasnoun{megalios2005},
which are described in the following.  

\subsection{X-ray method}
\label{sec:xray}
\begin{figure}
\centering
\includegraphics[width=0.4\linewidth]{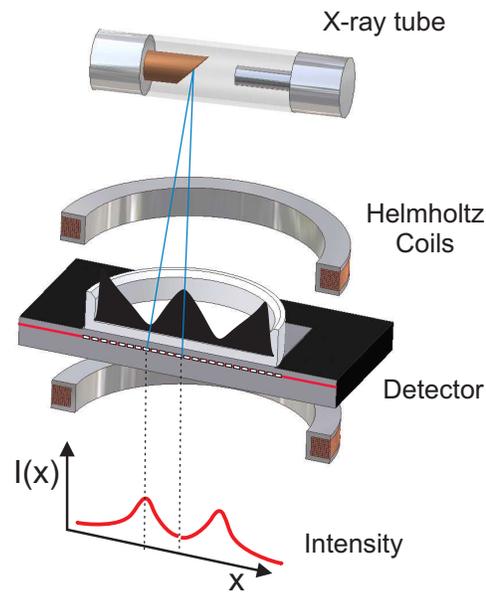}
\caption{The experimental setup for the X-ray method by \protect\citeasnoun{richter2001}}
\label{fig:xraysetup}. 
\end{figure}

\begin{figure}
\centering
\includegraphics[width=0.7\linewidth]{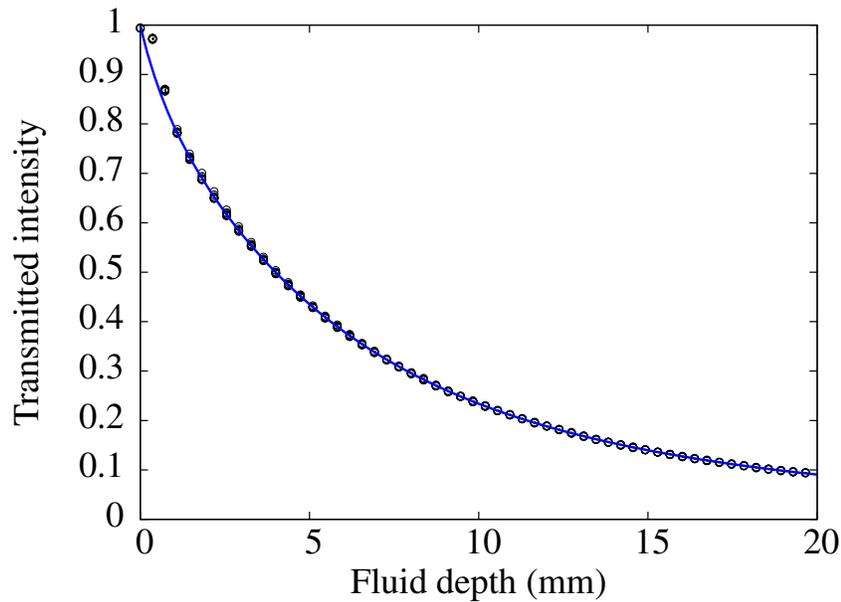}
\caption{The transmitted intensity as a function of the fluid depth. The solid
line is a fit with equation~\eref{eq:expcurve}.}
\label{fig:wedge}
\end{figure}

The X-ray apparatus comprises a stable X-ray point source, that emits radiation
vertically from above through the fluid layer. The container is placed midway
between a water cooled Helmholtz pair of coils, which generate a DC magnetic field of up to $40\,\mathrm{mT}$.
Directly below the container an X-ray camera with $512\times512$ pixels is located, which
measures the transmitted intensity at every pixel in one plane underneath the
fluid~\cite{richter2001}. This setup is depicted in figure~\ref{fig:xraysetup}.
The transmitted intensity of the X-rays is directly related to the height of the
fluid above every corresponding pixel. To calibrate this relation,
we use a wedge of known size, fill it with ferrofluid and place it in the empty container.
In this calibration image, we therefore know the height of the fluid. 
Figure~\ref{fig:wedge} shows the calibration data from the wedge.
These are then fitted with an overlay of three
exponential functions
\begin{equation}
I(h)=I_0\cdot\sum\limits_{i=1}^3 \alpha_i\exp(-\beta_i h)
\label{eq:expcurve}
\end{equation}
as a practical approximation, denoted by the solid line in figure~\ref{fig:wedge}.
Further details can be found in \citeasnoun{gollwitzer2007}. 

\begin{figure}
\centering
\includegraphics[width=0.7\linewidth]{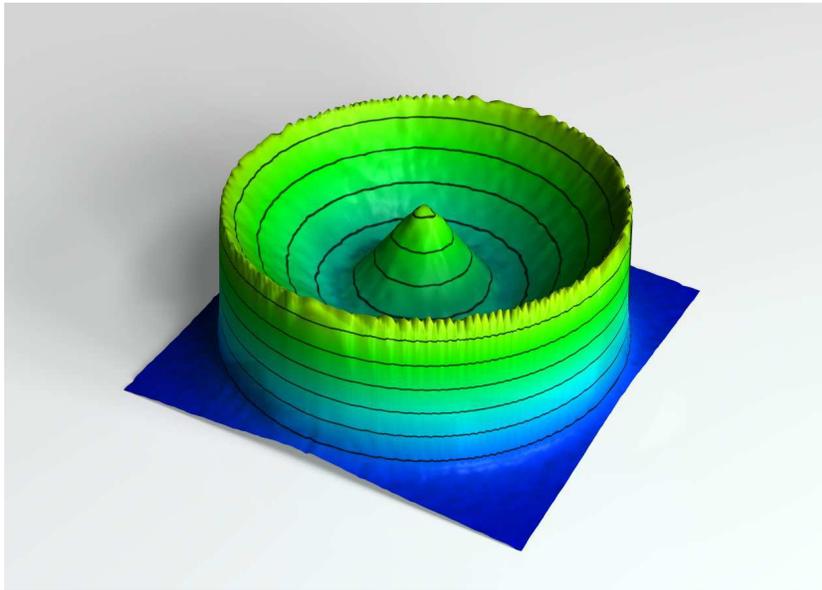}
\caption{Three-dimensional reconstruction of the surface of the single spike at
an induction of $B=22.69\,\mathrm{mT}$. A related movie can be accessed \href{http://www.staff.uni-bayreuth.de/~btp916/IKYDA/}{here.}}
\label{fig:3d}
\end{figure}
After applying the inverse of \eref{eq:expcurve} to an arbitrary image from the detector, we finally end up
with a complete three-dimensional surface topography of the filled container. A reconstruction for one
specific field is displayed in figure~\ref{fig:3d}. A survey of the surface topography for different 
fields is provided by the \href{http://www.staff.uni-bayreuth.de/~btp916/IKYDA/}{related movie.}

\subsection{Laser method}
\begin{figure}
\centering
\includegraphics[width=0.6\linewidth]{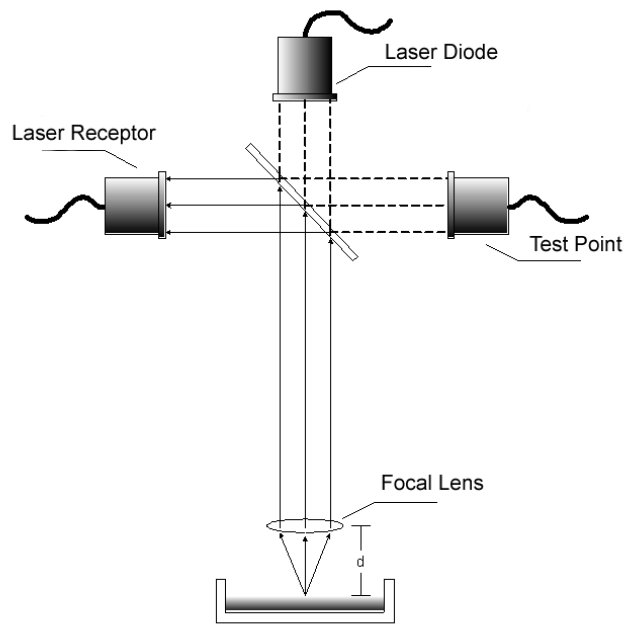}
\caption{The experimental setup of the laser method by \protect\citeasnoun{megalios2005}. The solid lines denote the path of the light reflected on the surface. The path sketched by the dashed lines is used to 
verify that the laser operates. }
\label{fig:lasersetup}
\end{figure}

The laser method developed by \citeasnoun{megalios2005} enables precise,
relative measurements of the extrema of the surface topography. 
Figure~\ref{fig:lasersetup} depicts the principle of operation. The container with
the ferrofluid is situated in a long solenoid, which generates a vertical magnetic field. 
The solenoid is $33\,\mathrm{cm}$ long with an internal diameter of $13\,\mathrm{cm}$ and
an external diameter of $14\,\mathrm{cm}$. It has $1124$ windings and produces up to
$21\,\mathrm{mT}$ at its centre, with a variation of less than $1\,\%$ at
the experimental region. 

A laser beam is directed at the fluid surface through a semitransparent
mirror, which splits the beam into two parts. One part is deflected sideways onto
a test point and serves as an indicator, whether the laser operates correctly (the dashed path 
in figure~\ref{fig:lasersetup}). The other beam is focused on the fluid surface 
and the reflected light is deflected by the semitransparent mirror onto a photodiode detector (the solid path 
in figure~\ref{fig:lasersetup}). 

The position of the focal spot can be adjusted by
means of a micrometre screw. The maximum of the reflected 
intensity is reached, when the direction of the beam coincides with the normal vector of the
surface at the focus spot and the distance of the lens from the surface 
is equal to the focal length. In normal operation, the beam is oriented
vertically - thus the intensity of the signal reaches its maximum when the focus
spot hits an extreme point of the surface, namely the top of a spike or the
minimum in the centre of the meniscus. By tracing the maximum intensity of the
reflected beam and recording the position of the laser optics, we get the
position of the spike with micrometre resolution relative to some reference point.
Also the absolute height of the spike above the bottom of the container can be determined by
setting the reference point at the top of the container edge. 

\section{Governing equations and computational analysis}
\begin{figure}
\centering
\includegraphics[width=0.4\linewidth]{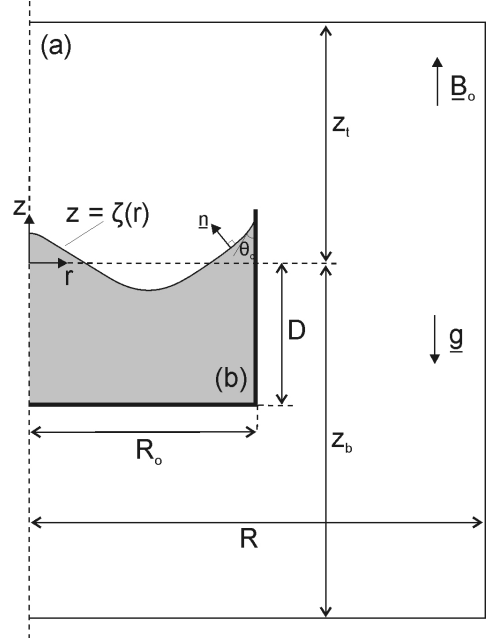}
\caption{A small axisymmetric ferrofluid pool in the magnetic field.}
\label{fig:fluidpool}
\end{figure}
A scheme of a small cylindrical ferrofluid pool in a vertical magnetic field is
shown in figure~\ref{fig:fluidpool}. The surrounding air and the embedded ferrofluid are denoted by
(a) and (b), respectively. The applied field can be produced either by
a pair of Helmholtz coils or a solenoid of suitable dimensions. It is
uniform i.\,e., of constant strength and vertical orientation, in a
region far away from the pool. The field uniformity, however, is
disturbed in the neighborhood of the pool, due to the demagnetizing
field of the pool itself. Therefore, the applied magnetic field could
be taken uniform only away from the pool. In the following, the magnetic field
distribution and the free surface deformation are taken as axially symmetric about 
the $r = 0$ axis (cf. figure~\ref{fig:fluidpool}).

The field distribution in regions (a) and (b) is governed by the
equations of magnetostatics. The Gauss law for the magnetization
reads
\begin{equation}
\nabla\cdot\underline B = 0,
\label{eq:gausslaw}
\end{equation}
where $\underline B$ is the magnetic induction. Since the
magnetic field $\underline H$ is irrotational it can
be derived from a magnetostatic potential $\underline H \equiv \nabla u$
both inside and outside the ferrofluid and, provided
that the materials are isotropic, it is parallel to $\underline B$
and so is the magnetization 
\begin{equation}
\underline B = \mu\underline H = \mu_0(\underline H+\underline M).
\label{eq:bhm}
\end{equation}
The magnetic permeability $\mu$ is constant in non-magnetic media, namely
$\mu_\text{a}=\mu_0 = 4\pi\times 10^{-7}\mathrm{H/m}$; inside
the ferrofluid, it depends on the field. Two different constitutive
equations are used to account for the field dependence on the magnetization. The
first one comes from Langevin's theory for monodisperse colloidal suspensions
(equation~\ref{eq:langevin}). The second one comes from a polydisperse model by
\citeasnoun{ivanov2001} that is based on the assumption of a gamma distribution of particle diameters.

Writing equation~\eref{eq:gausslaw} in terms of the magnetostatic potential, $u$, and
taking into account the equation~\eref{eq:bhm} yields
\renewcommand{\theequation}{\arabic{equation}a,b}
\begin{equation}
\nabla^2 u_\text{a} =0, \quad \nabla\cdot(\mu\nabla u_\text{b})=0
\end{equation}
\renewcommand{\theequation}{\arabic{equation}}
\label{eq:potential}
inside the non-magnetic phase~(a) and inside the magnetic phase~(b),
respectively.

Equlibrium is governed by force balance along the ferrofluid free
surface which is stated by the magnetically augmented Young-Laplace
equation of capillarity
\begin{equation}
-g\Delta\rho \zeta + \frac{1}{2}\mu_0\int\limits_0^{H_\text{bs}} M(H')\mathrm{d}H' 
+ 2\aleph\sigma = K, \text{at } z=\zeta(r),\quad 0\leq r\leq R_0
\label{eq:younglaplace}
\end{equation}
where $g$ is the gravitational acceleration, $\rho$ is
the density, $\sigma$ is the surface tension and $\zeta$ is the vertical
displacement of the free surface parametrized by the radial coordinate $r$,
i.e., $\zeta = \zeta(r)$. The upper limit $H_\text{bs}$ of the integral in the
magnetization term is the field strength in the ferrofluid, evaluated at the free
surface, i.e. at $z = \zeta(r).$

The reference pressure $K$ is constant at the free
surface. The unit normal to the free surface $\underline n$ 
and the local mean curvature of the free surface $2\aleph$ are 
\renewcommand{\theequation}{\arabic{equation}a,b}
\begin{equation}
\underline n = \frac{-\zeta_r\underline e_r + \underline e_z}{\sqrt{1 + \zeta_r^2}},
\quad 2\aleph =
\frac{1}{r}\frac{\mathrm{d}}{\mathrm{d}r}\left(\frac{r\zeta_r}{\sqrt{1+\zeta_r^2}}\right),
\end{equation}
\renewcommand{\theequation}{\arabic{equation}}
\label{eq:diffgeo}
where $\underline e_r$ and $\underline e_z$ are mutually orthogonal unit vectors
along the $r$- and $z$-axis, respectively, and $\zeta_r \equiv
\mathrm{d}\zeta/\mathrm{d}r$. 

The reference pressure $K$ in equation \eref{eq:younglaplace} is determined by the constraint, that the
ferrofluid volume is of fixed amount 
\begin{equation}
2\pi\int\limits_0^{R_0} \zeta r\mathrm{d}r =C=\text{const},
\label{eq:volumeconstraint}
\end{equation}
i.e. we assume an incompressible liquid. The coordinate system, i.e. the location of the
$z = 0$ line, is chosen such that $C=0$.

The set of the governing equations \eref{eq:potential}, \eref{eq:younglaplace}
and \eref{eq:volumeconstraint}, needs to be solved
for the magnetostatic potential $u_a(r, z)$ and
$u_b(r, z)$, the free surface shape $\zeta(r)$ and the
reference pressure $K$, taking into account the following boundary
conditions (see also figure~\ref{fig:fluidpool}):
\seteqabctag{a,b}
\begin{equation}
\frac{\partial u_\text{a}}{\partial r} = \frac{\partial u_\text{b}}{\partial r} = 0, \quad \zeta_r=0  \text{ at } r=0 
\renewcommand{\theequation}{\arabic{equation}}
\label{eq:axisym}
\end{equation}
\seteqabctag{a,b}
\begin{equation}
u_\text{a}=u_\text{b},\quad \mu\underline n\cdot \nabla u_\text{b} = \mu_0\underline n\cdot \nabla u_\text{a}
\quad \text{ at } 
z=\zeta(r) \text{ and } 0\leq r\leq R_0 
\renewcommand{\theequation}{\arabic{equation}}
\label{eq:potcont}
\end{equation}
\addtocounter{equation}{-1}
\seteqabctag{c,d}
\begin{equation}
u_\text{a}=u_\text{b},\quad \mu\frac{\partial u_\text{b}}{\partial r} = \mu_0
\frac{\partial u_\text{a}}{\partial r} 
\quad\text{ at } r=R_0 \text{ and } -D\leq z \leq \zeta(R_0) 
\end{equation}
\addtocounter{equation}{-1}
\seteqabctag{e,f}
\begin{equation}
u_\text{a}=u_\text{b},\quad \mu\frac{\partial u_\text{b}}{\partial z} = \mu_0
\frac{\partial u_\text{a}}{\partial z} 
\quad\text{ at } z=-D, 0\leq r\leq R_0
\end{equation}
\seteqabctag{}
\begin{equation}
u_\text{a} = 0  \text{ at } z=-z_\text{b} 
\label{eq:magpotfixed}
\end{equation}
\seteqabctag{a,b}
\begin{equation}
\frac{\partial u_\text{a}}{\partial z} = \frac{B_0}{\mu_0} \text{ at } z=z_t, \quad
\frac{\partial u_\text{a}}{\partial r} = 0 \text{ at } r=R
\label{eq:uniformfield}
\end{equation}
\seteqabctag{}
\begin{equation}
\zeta_r = \cot \theta_c \text{ at } r=R_0.
\label{eq:contactangle}
\end{equation}
\seteqabctag{}
The subscripts $r$ and $z$ denote differentiation with respect to $r$ and $z$,
respectively. Equations~\eref{eq:axisym} are the conditions that the shape of the free
surface and the magnetostatic potential are axially symmetric. Equations~\eref{eq:potcont}
are statements of the continuity of the potential and of the normal
component of the magnetic induction across interfaces between two media
with different magnetic permeabilities. A datum for the potential is
fixed by \eref{eq:magpotfixed}. Equations~\eref{eq:uniformfield} are the conditions that the magnetic
field be uniform far away from the pool. A contact angle
$\theta_c$ is prescribed by equation \eref{eq:contactangle} and 
reflects the wetting properties of the magnetic liquid in contact with the
solid wall of the container.

The governing equations give rise to a nonlinear, free boundary problem,
owing to the presence of the free surface, the location of which enters
the equations nonlinearly and is unknown a priori. An additional
nonlinearity comes from the constitutive equation for the magnetization of the fluid.
Such a problem is only amenable to
computer-aided solution methods. The choice is the combination
of Galerkin's method of weighted residuals and finite element basis
functions~\cite{kistler1983}.
\begin{figure}
\centering
\includegraphics[width=0.4\linewidth]{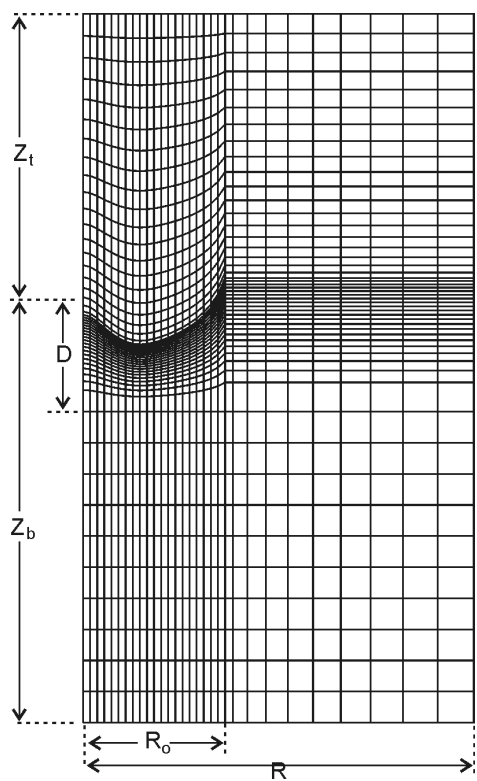}
\caption{Sample of the discretized domain}
\label{fig:mesh}
\end{figure}
Here we will only outline the application of the method. Details
can be found in previous works by \citeasnoun{boudouvis1988} and \citeasnoun{papathanasiou1998}. 
The domain is tessellated into
nine-node quadrilateral elements between vertical spines and
transverse curves whose intersections with each spine are located at
distances that are proportional to the displacement of the interface
along that spine. The tessellation creates a mesh of nodes and at each
node a finite element basis function is assigned that is unity at
that node and zero at all other nodes. As the basis functions, we choose a
quadratic polynomials of the independent variables $r$ and $z$. 
A sample of the computational
mesh is shown in Figure~\ref{fig:mesh}. The dependent variables
$u_\text{a}(r,z), u_\text{b}(r,z)$ and $\zeta(r)$ are
approximated in terms of a truncated set of the finite element basis
functions. The governing equations are reduced with Galerkin's method
to a set of nonlinear algebraic equations for the values of the
unknowns $u_\text{a}, u_\text{b}$ and $\zeta$ at the nodes and
for the value of $K$.

At fixed values of the physical parameters, the nonlinear algebraic
equation set is solved by Newton iteration. The parameter of interest
here is the applied magnetic induction $B_0$, which appears
in the boundary conditions (cf. equation~\ref{eq:uniformfield}a). Solution families, i.e.,
solutions at sequences of different values of $B_0$ are
systematically traced with first-order continuation. 
The computational results reported are obtained with a mesh of 24000 nodes. 
The sensitivity of the spike height to further mesh refinement is practically 
negligible; namely, less than 0.2\,\% when doubling the mesh density. 
Three to five Newton steps are needed for the convergence at each value
of the continuation parameter.

\section{Comparison of experimental and numerical results}
\begin{figure}
\centering
\includegraphics[width=0.7\linewidth]{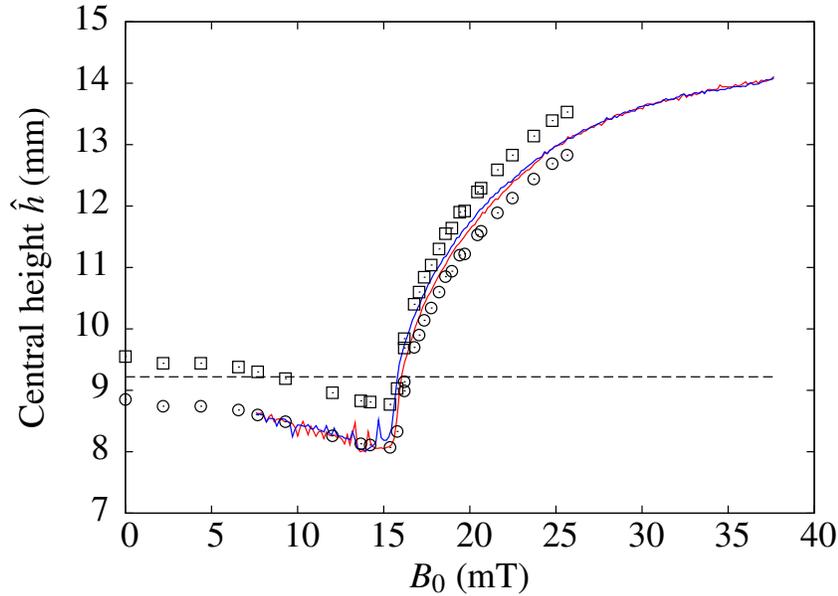}
\caption{Height of the single spike or the minimum in the centre, respectively.
The numbers given are the total absolute height of the fluid above the bottom of the
container. For clarity, the individual 200 data points for one sweep
are connected by a solid line. The red (blue) line denotes the values from the X-ray measurement for
increasing (decreasing) field, respectively. 
The open squares show the result of the laser
measurement. The open circles denote the laser data shifted on top of the X-ray
data. The dashed line represents the filling level of the fluid according to the
weight, neglecting the effects of the meniscus.}
\label{fig:expdata}
\end{figure}

We record the surface profile for $200$ different magnetic inductions 
$B_0 \in \left[7.6\,\mathrm{mT}\dots37.7\,\mathrm{mT}\right]$ with the X-ray method described above. 
Unevitably, the magnetic induction in the neighborhood of the container is
distorted and emphasized in comparison to the induction for empty coils.  
The given values $B_0$ denote the spatially averaged
magnetic induction below the container at a vertical distance of $23.8\,\mathrm{mm}$
from the bottom of the fluid layer. This corresponds to the boundary of the
computational domain in the calculations.
The height and position of the 
extreme point of the surface topography in the centre has been determined by fitting a paraboloid to a
small circular region of the surface 
with a diameter of $1.5\,\mathrm{mm}$. 
Figure~\ref{fig:expdata} shows the resulting central height $\hat h(B_0)$. 
The red solid line marks the data for increasing induction. 
The magnetic fluid first rises at the edge of the container, 
thus the level of
fluid in the centre of the vessel drops. The central height then corresponds to the
minimum level in the centre. At a magnetic induction of around
$16\,\mathrm{mT}$, a single spike emerges in the centre that continues to rise
for increasing induction. The central height then corresponds to the height of
this spike. A small hysteresis is found when decreasing the field again. See the blue line
in figure~\ref{fig:expdata}.

Using the laser method, we performed measurements of the spike height for $29$ different magnetic
inductions from $0$ to $25.65\,\mathrm{mT}$, which are also plotted in figure~\ref{fig:expdata}. 
By focusing the laser beam on top of the container edge, a reference point was
taken to get absolute values for the central height $\hat h$ above the bottom of the
container, denoted by the open squares. They differ from the X-ray measurement by a shift of
$0.7\,\mathrm{mm}$. However, if the reference point is adjusted by this shift,
we find a nice coincidence of the data points from both methods, as shown by the
open circles in figure~\ref{fig:expdata}.

The inaccuracy of the reference point of the laser method can well be explained
from the fact, that the laser beam cannot be focused precisely onto the top
edge of the container. The vessel is machined from aluminum with a quite rough
surface and diffuses the incident
laser beam, which leads to the observed shift. This has been experimentally verified 
by comparing the height measurements of the bare aluminium and a ferrofluid surface
at the same level. The difference 
in the reading is large enough to explain the shift between the laser data and the X-ray data.
On the other hand, the accuracy of the reference point of the X-ray measurement
depends on the positioning of the calibration wedge. The resolution of this
position is limited by the lateral resolution of the detector, which leads to an
estimate of the absolute error of $0.2\,\mathrm{mm}$. Due to the roughness of the aluminium vessel,  
the X-ray data seem to be more precise than the laser data concerning the absolute
height in the present experiment. Relatively, both yield practically the same result. 

Further deviations may stem from the different ambient temperature at which the data were
taken. Whereas in Bayreuth, the lab temperature was stabilized at $21\,\pm1$\,\textdegree
C, the temperature in Athens was $30\,$\textdegree C. This leads to a reduced
magnetisation and may be the origin of a reduced spike height for higher fields (cf.
figure~\ref{fig:expdata}). 


\begin{figure}
\centering
\includegraphics[width=0.5\linewidth]{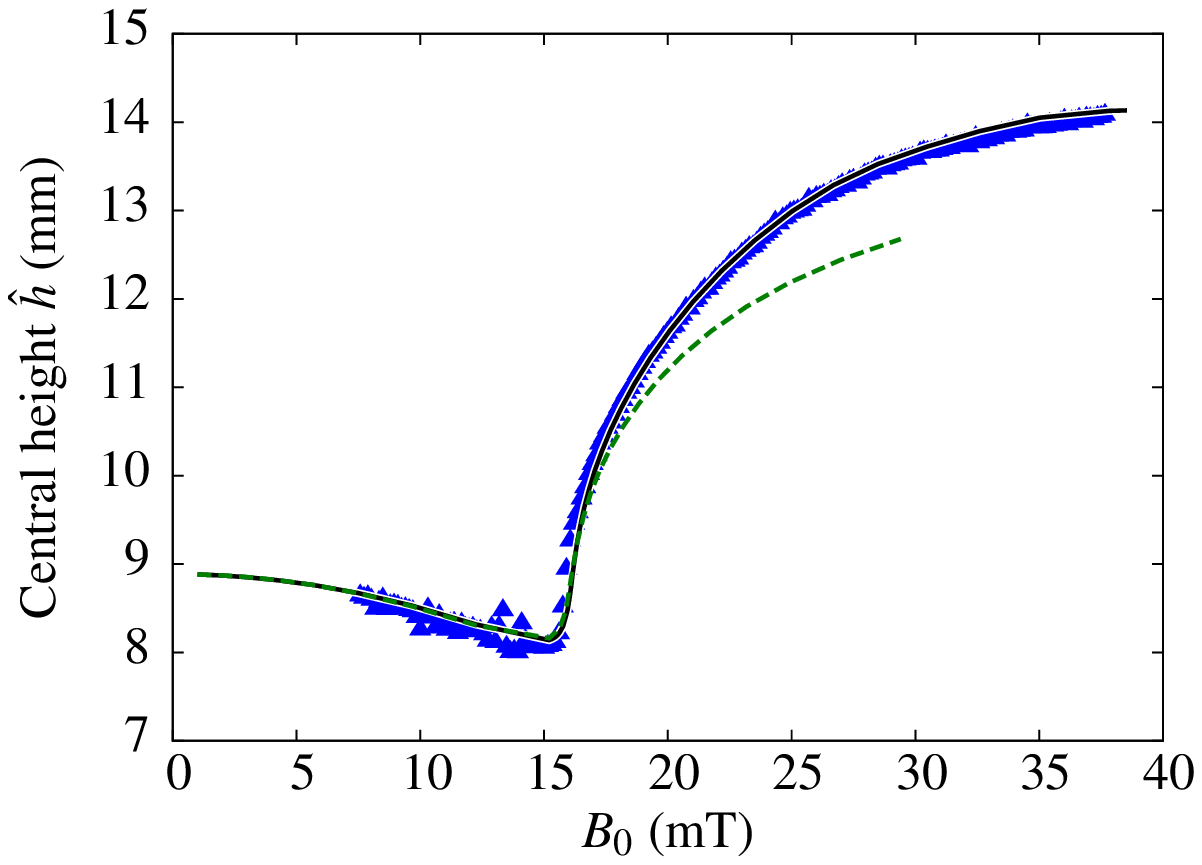}\\
(a)\\
\includegraphics[width=0.5\linewidth]{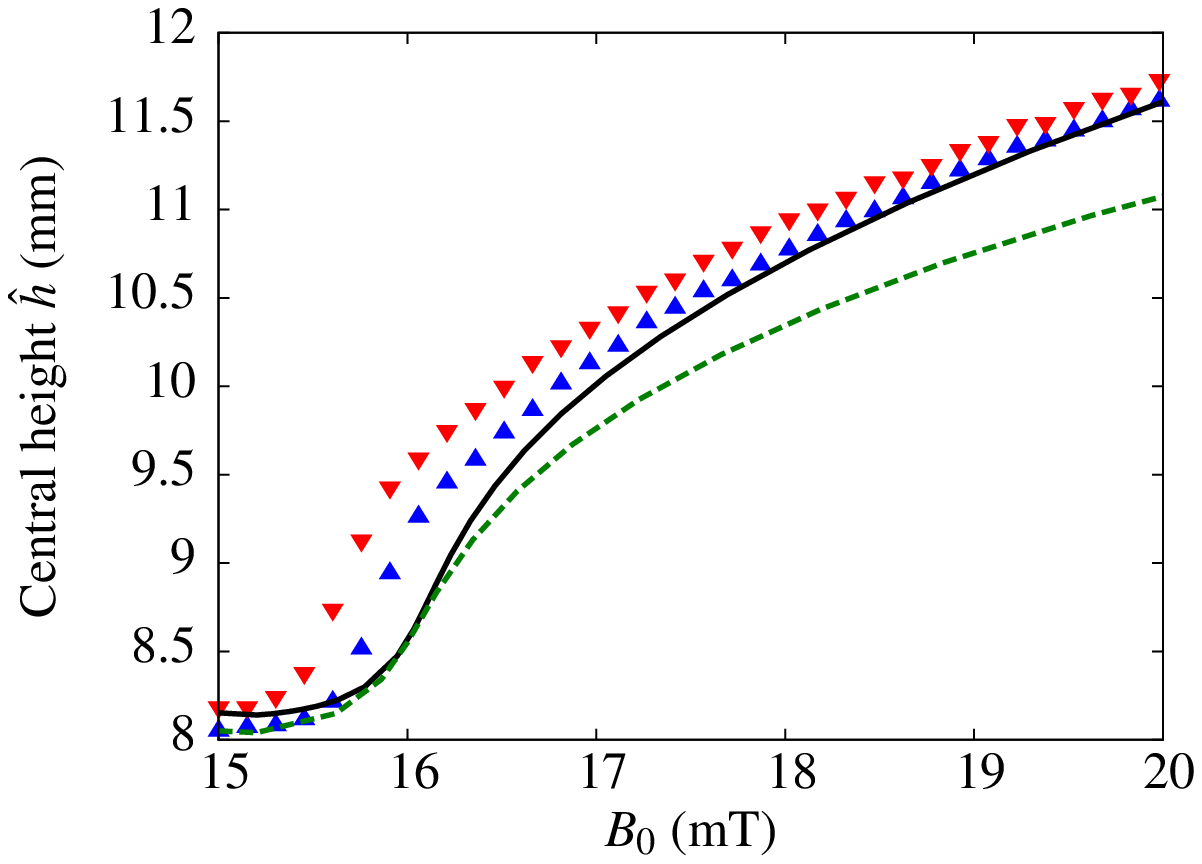}\\
(b)\\
\caption{The central height $\hat h$ from the computations (lines) and from the 
X-ray experiment (triangles). Blue upward (red downward) triangles denote the values for
increasing (decreasing) induction, respectively.
The dashed green line is based on Langevin's law for the
magnetisation, while the black solid line employs \protect\possessivecite{ivanov2001} model.
(a)~Full range (b)~Zoom}
\label{fig:allcompare}
\end{figure}

After successfully comparing the results of the two measurement techniques, we
now present the numerical predictions. 
The results of the computational analysis are depicted in
figure~\ref{fig:allcompare} together with the X-ray data. The value
corresponding to the central height $\hat h$ of the measurements is the height 
at the axis of symmetry $\left.h\right|_{r=0}=\left.\zeta\right|_{r=0}+D$, where $D$ denotes the filling level. Two computational
equilibrium paths are shown for two different models for the magnetziation
$M(H)$. The green dashed line displays the numerical result using Langevin's
equation~\eref{eq:langevin}, while the black line makes use of the model by
\citeasnoun{ivanov2001}. For magnetic inductions up to $17\,\mathrm{mT}$ there is no
noticable difference between both results. This is explained by the coincidence of
both magnetization laws up to an internal field of $H\approx10\,\mathrm{kA/m}$, as
shown in figure~\ref{fig:magcurve}. For higher fields, however, Langevin's law
is no longer valid. This leads to a rather big deviation between both
theoretical curves. Regardless of the underlying magnetisation curve, the numerical
solutions show a continuous behaviour of $\hat h$ in the full range of $B_0$. In particular, no
turning points are traced on the curve. This is mentioned, since a pair of turning points,
if existed, should imply a hysteresis in surface deformation observed when increasing and
then decreasing $B_0$. Thus, from the theoretical calculations we do not expect any hysteresis. 
We stress this fact, because in the case of an infinitely extended container, a hysteresis is both expected
theoretically \citeaffixed[e.g.]{friedrichs2001}{cf.} and found in the
experiment and numerical calculations~\citeaffixed[e.g.]{gollwitzer2007}{}.
Moreover, we tested the influence of the contact angle. A computation for
$\theta_c=20\,$\textdegree  does not deviate more
than $60\,\upmu\mathrm{m}$ from the above calculation over the full range of $B$. Thus,
the spike height does only weakly depend on the contact angle. 

Next, we compare the measurements with the numerical results. 
In the full range, the experimental data coincide very well with the more
advanced computations taking into account the magnetization law by \citeasnoun{ivanov2001}. 
The difference is within only $1\,\%$ of the absolute height, except near the threshold, where 
it amounts to $6\,\%$. This is natural for a sigmoidal function, where close to the steep
increase the error can become arbitrarily large, when there are unertainties of the
control parameter. The difference between the thresholds in theory and experiment amounts to
at most $3\,\%$ as can be seen from Figure~\ref{fig:allcompare}\,(b). It shows an enlarged view 
of $\hat h$ in the immediate vicinity of the threshold.
In opposite to the numerical results, we observe
in the experiment a small hysteresis between the data for increasing $B_0$, as marked
by upwards triangles, and the one for decreasing $B_0$, denoted by downward
triangles. The hysteresis is in the range of $\Delta B_0=0.2\,$mT. The origin for this hysteresis is a priori not clear.

Note, that \citeasnoun{gollwitzer2007}, e.g.,
measure a hysteresis of $\Delta B^\mathrm{\infty}=0.17\,$mT for the same
ferrofluid as in our case in a container with a diameter of $\approx 10 \times
\lambda_c$. Remarkably, this value is in the same range as the one observed
above.  If the hysteresis in our experiment would be of the same nature, it
should be much smaller due to the imperfection caused by the container
edge~\cite{cross1993}. Therefore we suspect another mechanism. One candidate is
a hysteretical wetting of the cylindrical wall. The difference between the advancing and the
receding contact angle can be as large as $10\,$\textdegree\ for a surface that has not been specially prepared 
\cite{degennes1985,dussan1979}. Moving the contact line always costs energy.
This may explain the small hysteresis of the spike height for increasing and decreasing induction. 
In our experiment, this effect is important, because the
interfacial area between the fluid and the vertical wall is comparable to the free surface of the
fluid.

\begin{figure}
\begin{minipage}{0.45\linewidth}
\centering
\includegraphics[width=\linewidth]{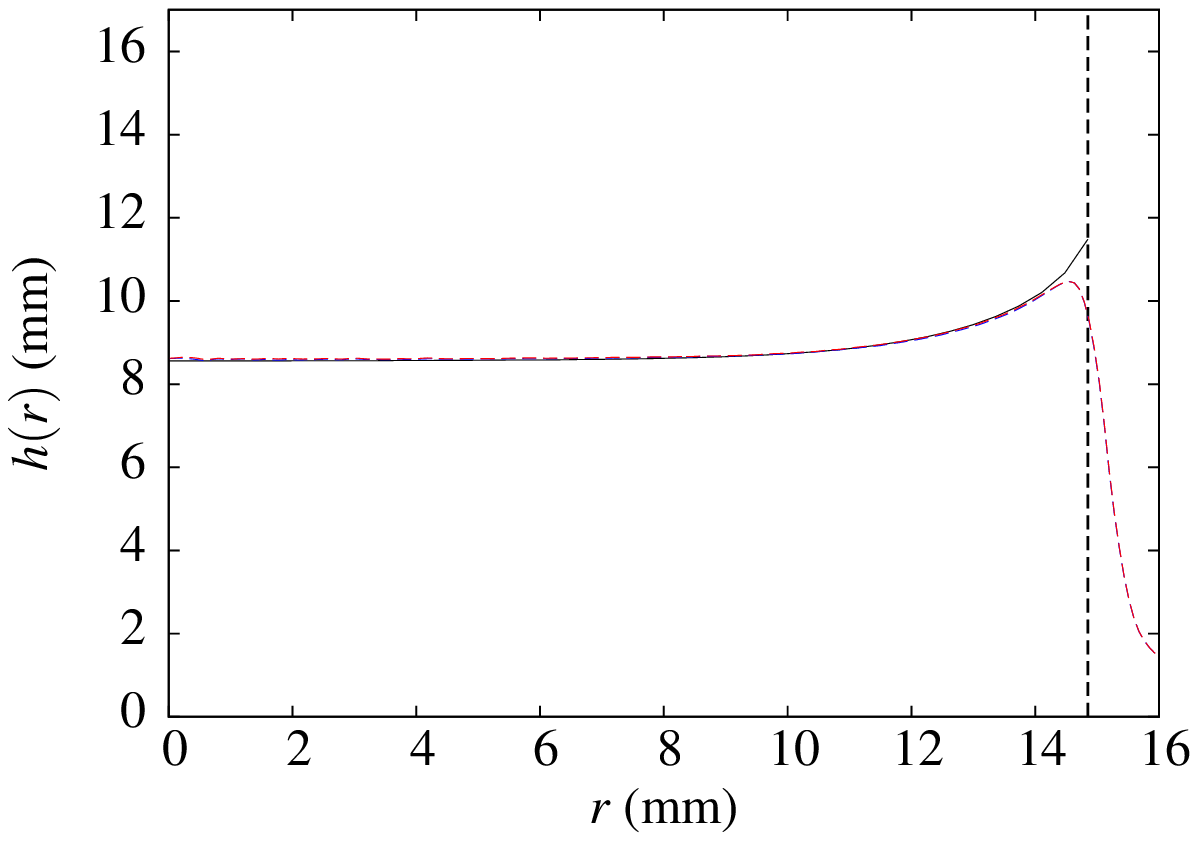}\\
(a)
\end{minipage}\hfill
\begin{minipage}{0.45\linewidth}
\centering
\includegraphics[width=\linewidth]{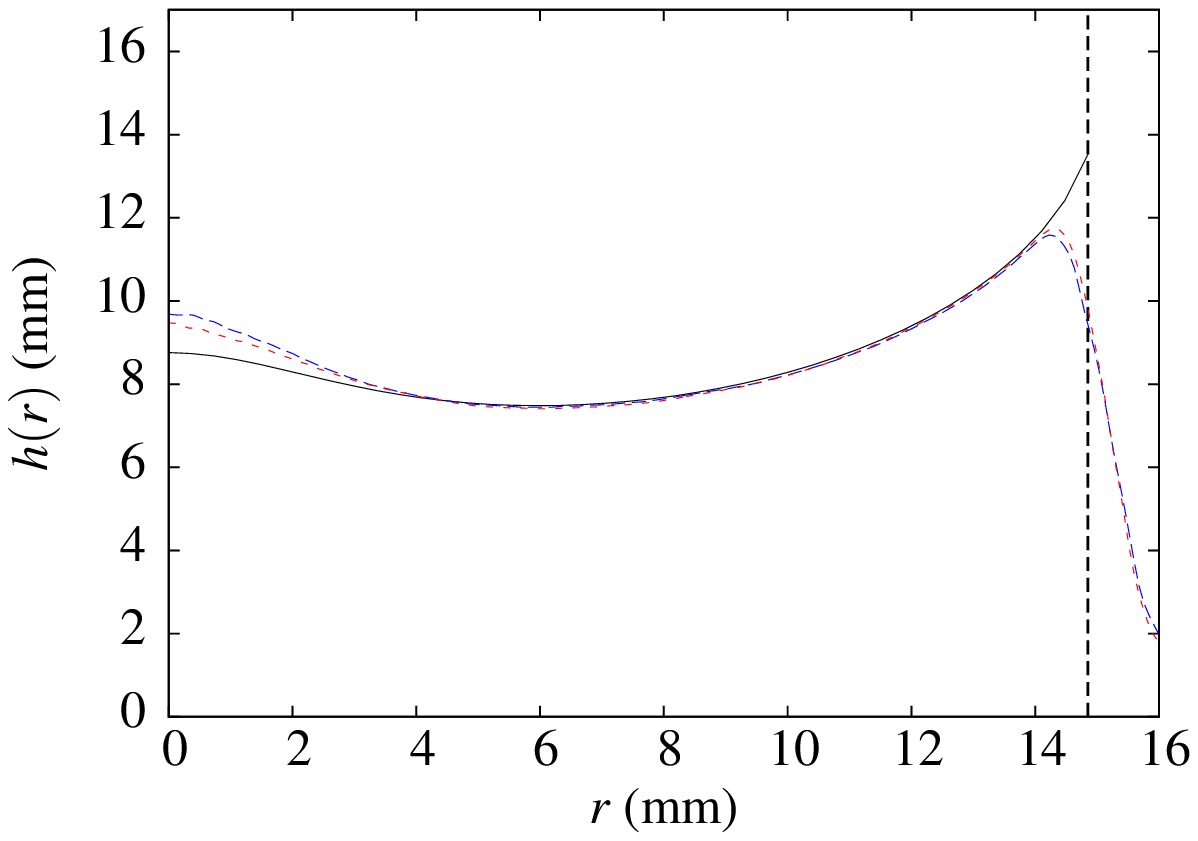}\\
(b)
\end{minipage}\\
\begin{minipage}{0.45\linewidth}
\centering
\includegraphics[width=\linewidth]{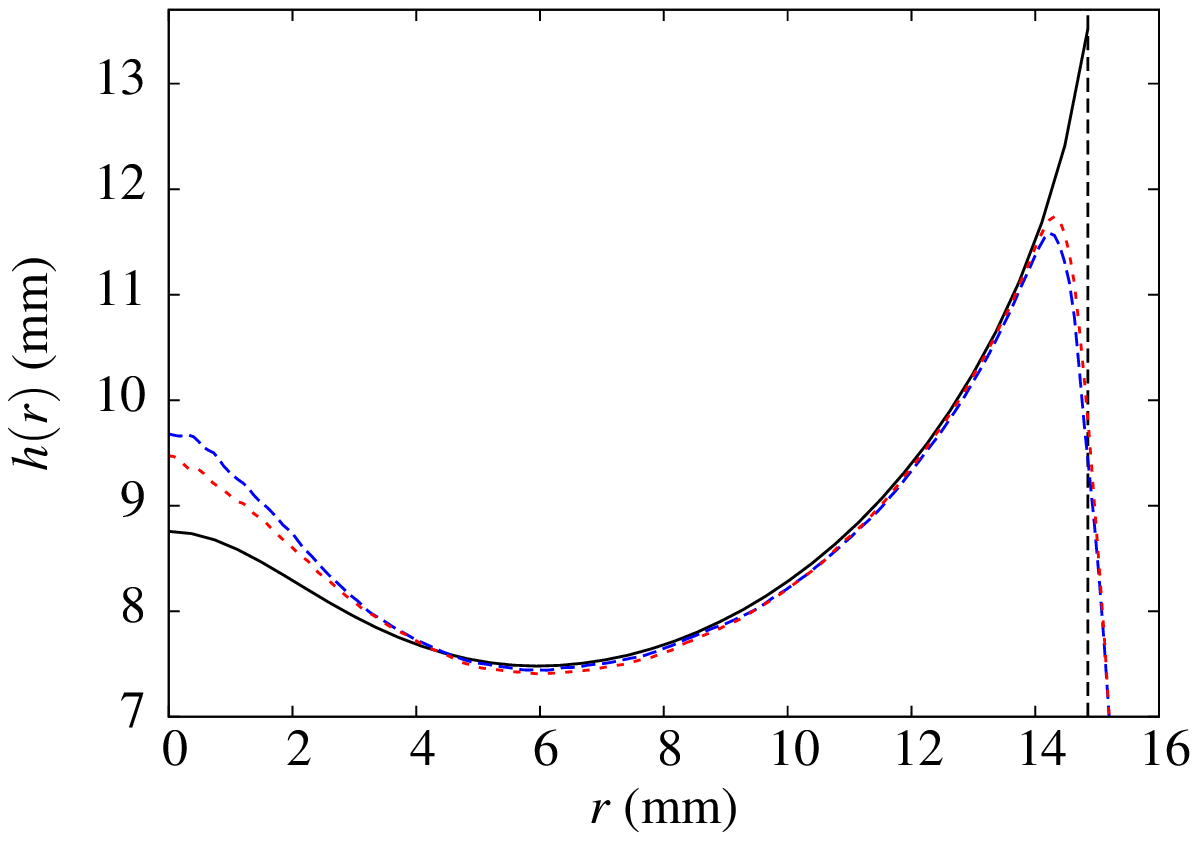}\\
(c)
\end{minipage}\hfill
\begin{minipage}{0.45\linewidth}
\centering
\includegraphics[width=\linewidth]{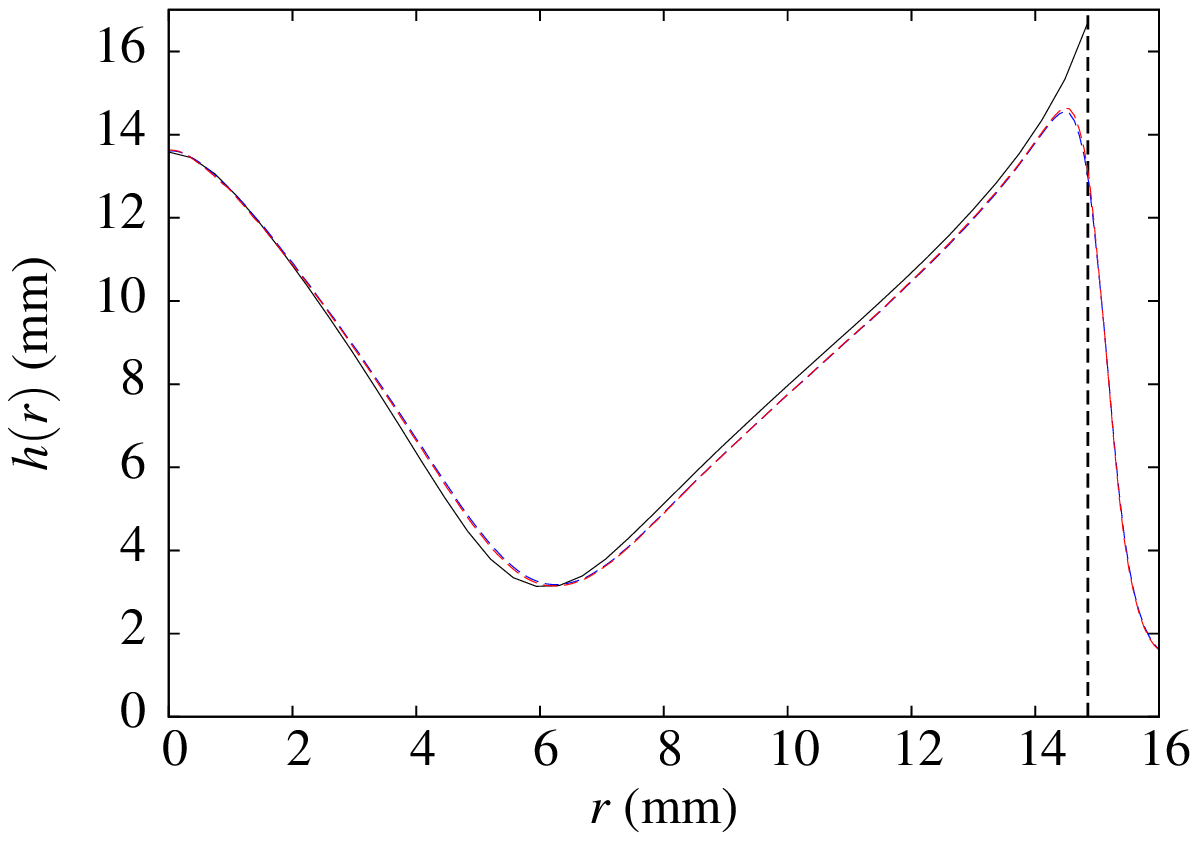}\\
(d)
\end{minipage}\\
\caption{The ferrofluid free surface shapes at various magnetic field strengths. 
The solid lines show the numerically computed profiles, while the red (blue) dashed line
gives radial interpolation of the experimental data for increasing (decreasing) magnetic induction. 
The vertical dashed line shows the side wall of the container.
(a)~$B=7.39\,\mathrm{mT}$, (b)~$B=16.17\,\mathrm{mT}$, (c) Zoom of (b), (d)~$B=29.70\,\mathrm{mT}$ }
\label{fig:radprof}
\end{figure}

The availability of the complete surface topography from the X-ray method 
allows us not only to compare the central height, but also the shape of the
spike or meniscus, respectively. Examples for the free surface shapes at selected values of
the magnetic induction are shown in figure~\ref{fig:radprof}. The experimental
data have been averaged angularly around the centre of the observed spike, which
is not exactly in the centre of the container in the experiment. This off-centre
distance is rather small ($0.05\,\mathrm{mm}$), however it must be taken into
account, otherwise the averaging would disturb the shape of the spike.

Similiarly to the comparison of the height alone, there are only slight
differences between the computations and the experimentally observed shape.
Most notably, we discern a drop at the edge of the container. The reason for this difference is
two-fold: first, the angular average does not work well near the container
border, because the centre of the spike is off-axis, as explained before.
Second, the X-ray method has problems to accurately detect the height near the
border, where the container wall shadows the X-rays. 
Further, for magnetic inductions near the threshold
(cf. figure~\ref{fig:radprof}\,(b) and (c)), the height of
the tip differs by $\approx 1\,\mathrm{mm}$. The reason is probably a slight
shift of the critical induction (cf. figure~\ref{fig:allcompare}), where the height is
very sensitive to small changes of the induction $B$.

The hysteresis, already observed from the central height alone, manifests
itself by a difference of the surface profiles for increasing and decreasing
induction.  Far away from the threshold, both profiles match nearly perfectly
(figure~\ref{fig:radprof}\,(a) and (d)), while there is a clear difference near
the threshold (figure~\ref{fig:radprof}\,(b) and (c)). The tip of the spike is
considerably smaller for an increasing magnetic induction, while the level of
the fluid near the container edge is higher. This corroborates   
a hysteretical wetting to be responsible for the hysteresis. 

Apart from these differences, the deviation between the computed and measured profiles is
around $1\,\%$. 

\section{Discussion and Conclusion} 
For a rotational symmetric system with broken up-down
symmetry we have reduced the container size until only a single entity remains. In the
case of the Rosensweig instability this is a single spike of ferrofluid. Whereas for our
fluid, the extended system exhibits a transcritical bifurcation to hexagons, here an imperfect bifurcation sets in,
and the axisymmetric free surface deformation evolves supercritically and monotonically.
This is at least
the outcome of the monolithic finite element approach, which takes into account the
side-wall effects, namely the wetting and the fringing field, as well as the
polydispersity of the fluid.  We find a convincing agreement between theory and two
independent measurement techniques, the errors being within $3\,\%$ without any adjustable parameter. 
The nonetheless observed hysteresis is due to the wetting.

Our findings immediately raise the issue of what is ``in between'', regarding the structure of the solution space -- 
that is surface deformation versus applied field and other key parameters -- as the size
of the pool grows laterally. This will be tackled in a forthcoming publication by
\citeasnoun{spyropoulos2006}.

A transition from a transcritical backward to an imperfect forward bifurcation under spatial 
constraints
has also been reported by \citeasnoun{peter2005shc}. Their spatial stripe forcing simultaneously breaks
the rotational and translational symmetry. In our case it is sufficient to
break the translational symmetry.

To conclude, we have quantitatively compared numerics and experiments of the Rosensweig
instability in a system of finite size. This is a specific example, how  external constraints may
change a perfect transcritical bifurcation to an imperfect supercritical one.

\section*{Acknowledgement}
We are grateful to Klaus Oetter for constructions and to Friedrich Busse, Werner Pesch,
Ingo Rehberg, and Uwe Thiele for useful discussions.
We thank Carola Lepski for measurements of the surface tension and Lutz Heymann
for determining the contact angle. 
Travel expenses for the scientific cooperation between the Greek and the German teams were
covered by the Greek `State Scholarships Foundation' (IKY) and the `German Academic
Exchange Service' (DAAD) through the IKYDA program. Specifically, this program allowed 
us to transport a
sealed ferrofluid container together with the related experimentators to perform X-ray
measurements in Bayreuth and Laser measurements in Athens.  
\section*{References}

\bibliographystyle{jphysicsB} 
\bibliography{alles_neu}

\end{document}